\documentclass[12pt]{iopart}

\expandafter\let\csname equation*\endcsname=\relax
\expandafter\let\csname endequation*\endcsname=\relax

\usepackage{comment} 
\usepackage{amsmath}
\usepackage{enumerate}
\usepackage{graphicx}
\usepackage{mathbbol}
\usepackage{amsfonts}

\newcommand{\sgn}{\operatorname{sgn}}

\newcommand{\vertt}{\,\vert\,}

\begin{document}
\title{ Asymmetric L\'evy flights  in the presence of absorbing boundaries} 

\author{Cl\'elia de Mulatier}
\address{CEA/Saclay, DEN/DM2S/SERMA/LTSD, 91191 Gif-sur-Yvette Cedex, France\\
Universit\'e Paris-Sud, LPTMS, CNRS (UMR 8626), 91405 Orsay Cedex, France}

\author{Alberto Rosso}
\address{Universit\'e Paris-Sud, LPTMS, CNRS (UMR 8626), 91405 Orsay Cedex, France}

\author{Gr\'egory Schehr}
\address{Universit\'e Paris-Sud, LPTMS, CNRS (UMR 8626), 91405 Orsay Cedex, France}

\begin{abstract}
We consider a one dimensional asymmetric random walk  whose jumps are identical, independent and drawn from a distribution  $\phi(\eta)$ displaying asymmetric power law tails  (i.e. $\phi(\eta) \sim c/\eta^{\alpha +1} $ for large positive jumps and  $\phi(\eta) \sim c/( \gamma |\eta|^{\alpha +1} )$ for large negative jumps, with $0 < \alpha < 2$). In absence of boundaries and after a large number of steps $n$, the probability density function (PDF) of the walker position, $x_n$, converges to an asymmetric L\'evy stable law of stability index $\alpha$ and skewness parameter $\beta=(\gamma-1)/(\gamma+1)$. In particular the right tail of this PDF decays as $c \, n/x_n^{1+\alpha}$.
 Much less is known when the walker is confined, or partially confined, in a region of the space. In this paper we first study  the case of a walker constrained to move on the positive semi-axis and absorbed once it changes sign. {In this case, the persistence exponent $\theta_+$, which characterizes the algebraic large time decay of the survival probability, can be computed exactly and we show that the tail of the PDF of the walker position decays as $c \, n/[(1-\theta_+) \, x_n^{1+\alpha}]$.}  This last result can be generalized in higher dimensions such as a planar L\'evy walker confined in a wedge with absorbing walls. Our results are corroborated by precise numerical simulations.
\end{abstract}

\maketitle

\section{Introduction and main results}

Let us consider a one-dimensional random walker, in discrete time, moving on a continuous line. Its position $x_n$ after $n$ steps evolves, for $n \geq 1$ according to  
\begin{eqnarray}\label{def_rw}
x_n = x_{n-1} + \eta_n \;,
\end{eqnarray}
starting from $x_0 = 0$. The random jumps variables $\eta_i$'s are independent and identically distributed (i.i.d.) according to a probability density function (PDF) $\phi(\eta)$ displaying (asymmetric) power law tails: 
\begin{eqnarray}\label{tail_jump}
\phi(\eta) \sim 
\begin{cases}
&\dfrac{c}{\eta^{1+\alpha}} \;, \; \eta \to +\infty \;, \vspace{2mm}\\
&\dfrac{c/\gamma}{|\eta|^{1+\alpha}} \;, \; \eta \to -\infty \;,
\end{cases}
\end{eqnarray}
where $\alpha$ is a positive number in the interval $(0,2)$. The tails display an asymmetry when $\gamma \neq 1$. In this case, the random walk is {\em Markovian}
and exhibits a super-diffusive behavior. Power-law distributions such as in
Eq.~(\ref{tail_jump}) have been initially studied in the early Sixties in economics~\cite{pareto} and in financial theory~\cite{mandelbrot2}. Later
on, these processes became very common in Physics, where they have found many applications, encompassing laser-cooling of cold
atoms~\cite{shlesinger_book}, random matrices~\cite{biroli,MSVV2013}, disordered systems~\cite{bouchaud}, photons in hot atomic vapours~\cite{mercadier},
and many others. One striking feature of such processes is that their statistical behavior is dominated by a few rare and very large events, whose occurrence is thus governed by the {\it tail} of the distribution.
Often the applications of L\'evy flights are restricted to the symmetric case when $\gamma=1$, however recently the asymmetric L\'evy flights have found applications in search problems \cite{koren1} and finance \cite{Stanley}. Diffusion in asymmetric disordered potential was recently considered in connection with the ratchet-effect \cite{dario}.

When the number of jumps $n$ is large, the PDF of the walker position $x_n$ exhibits a strong universal behavior, i.e. this PDF depends on very few characteristics of the initial jump distribution $\phi(\eta)$. For $\alpha >2$, only the bulk of the distribution matters through its average, $\mu = \langle \eta \rangle$ and variance $\sigma^2 = \langle \eta^2 \rangle- \langle \eta \rangle^2$.  On the other hand, for $1<\alpha <2$, the variance is not defined and the PDF depends on  $\mu$, but also on the tails. Hence it depends on $\alpha$, $c$ and $\gamma$.  To study the large $n$ behavior it is useful to write the walker position after $n$ steps in the scaling form~\cite{feller_book, hughes_book, metzler}:
\begin{eqnarray}\label{scaling}
x_n=  \mu \,  n  + y \, n^{1/\alpha}  \;.
\end{eqnarray}
  When $n \to \infty$, the fluctuations of the variable $y$ are described by a PDF which 
 is independent of  $n$ and of the details of $\phi(\eta)$, except for the index $\alpha$, the constant $c$ and the parameter $\gamma$, as mentioned above.

If we consider a free one-dimensional random walker (i.e. in absence of boundaries), we know, from the Central Limit Theorem, that  this PDF corresponds to the skewed $\alpha$-stable distribution, $R(y)$. This distribution is  conveniently defined by its characteristic function, $\psi(t) = \int_{-\infty}^{+\infty} \, d y \, R(y) e^{i y t} $:
\begin{align}\label{def1b}
\psi(t) &= 
\begin{cases}
&\exp \left[ -|a t|^\alpha \left(1 - {i}\,\beta\,\sgn(t) \tan(\pi \alpha/2)\right)    \right]  \qquad {\rm if} \; \alpha \neq 1\;, \\
&\exp \left[ -|a t| \left(1 + \frac{2i}{\pi}\,\beta\,\sgn(t)\ln{|t|}  \right)    \right] \qquad \; \; \; \;\;\;\; \;{\rm if} \; \alpha = 1\;,
\end{cases}
\end{align}
where $\alpha$ is the stability index, $\beta \in [-1,+1]$ is  the skewness parameter describing the asymmetry of $R(y)$ (i.e. the property that $\gamma \neq 1$), $a>0$ is the scale parameter  describing the width of the distribution, 
 and $\sgn(t)$ denotes the sign of $t$.  The PDF $R(y)$ admits  the exact asymptotic expansion (see for instance Ref.~\cite{hughes_book}):  
\begin{align}\label{free_levy}
R(y) & \underset{|y| \to \infty}{\sim} \frac{1}{\pi\, |y|} \sum_{k=1}^\infty \frac{  a^{\alpha k} \left(1 + \sgn(y) \beta\right)^k \sin{  \left(\frac{\alpha k \pi}{2} \right)} \,\Gamma(\alpha k +1)\,(-1)^{k+1} }{k!\,|y|^{\alpha k}}\;.
\end{align}
We observe that $R(y)$ inherits the power law tail $\propto |y|^{-\alpha -1}$ of the jump distribution $\phi(\eta)$~(\ref{tail_jump}), both when $y \to +\infty$ and $y \to -\infty$. One can further show that the amplitudes of the right and the left tails of $R(y)$ have exactly  the same value as the corresponding amplitudes of $\phi(\eta)$, namely $c$ and $c/\gamma$~\cite{hughes_book}. Thus from (\ref{free_levy}), the parameters $c$ and $\gamma$ can be related to $a$ and $\beta$ via
\begin{eqnarray}\label{c_and_gamma}
c = \frac{a^{\alpha} \, \sin{\left(\frac{\alpha \pi}{2}\right)\,\Gamma(\alpha +1)} }{\pi} \left(1+\beta\right) \qquad {\rm and} \qquad \gamma = \frac{1+\beta}{1-\beta}\;.
\end{eqnarray}

Much less is known in the presence of boundaries, which is the focus of the present paper. Here we will study the case $\mu = 0$. Hence the 
scaling variable describing the position of the walker after $n$ steps is simply~(\ref{scaling})  
\begin{eqnarray}\label{scaling2}
   y = \frac{x_n}{ n^{1/\alpha}}  \;.
\end{eqnarray}
As in the case without boundaries, we also expect that the PDF of $y$ is independent of  $n$ and of the details of $\phi(\eta)$ (except for  $\alpha$,  $c$ and $\gamma$). As a first example of a bounded domain, we consider a walker that has not changed sign up to time $t$.  An important property characterizing such random walks (\ref{def_rw}) is the survival probability, or the persistence \cite{SatyaReview, Bray}, defined as the probability that the walker, starting from $x_0 = 0$, is still alive after $n$ steps (having in mind that the walker ``dies" if its position changes its sign). Given the asymmetry of the jump distribution, one introduces two distinct survival probabilities $q_+(n)$ and $q_-(n)$ defined as
\begin{eqnarray}
q_+(n) = {\rm Prob.} [x_n\geq 0, \cdots , x_1 \geq 0 | x_0 = 0] \;, \label{q+}\\
q_-(n) =  {\rm Prob.} [x_n\leq 0, \cdots , x_1 \leq 0 | x_0 = 0] \label{q-} \;.
\end{eqnarray}
Of course for symmetric jump distribution $\phi(\eta) = \phi(-\eta)$, or equivalently for $\beta = 0$, one has $q_+(n) = q_-(n)$, but for asymmetric $\phi(\eta)$ as in (\ref{tail_jump}), one has $q_+(n) \neq q_-(n)$. For large $n$, one expects that $q_\pm (n)$ decay algebraically with two distinct persistence exponents $\theta_+ \neq \theta_-$
\begin{eqnarray}\label{asympt_S}
q_{+}(n) \underset{n \to \infty}{\propto} n^{-\theta_{+}} \qquad\qquad q_-(n) \underset{n \to \infty}{\propto} n^{-\theta_-} \;,
\end{eqnarray}
where the exponents $\theta{\pm}$ are expected to depend explicitly on $\alpha$ and $\beta$, $\theta_{\pm} \equiv \theta_{\pm}(\alpha, \beta)$. Even for symmetric jump distribution (i. e. $\beta = 0$), the computation of $\theta_\pm$ is not trivial, in particular because the method of image fails for L\'evy flights, due to the presence of non-local jumps \cite{klafter_image}. {In the asymmetric case, $\beta \neq 0$, the exponents $\theta_{\pm}$ have been studied
in the physics literature in Ref. \cite{koren1,koren2,dybiec}. Using a generalized version of the Sparre Andersen theorem, the persistence exponents $\theta_+$ and $\theta_-$ can be computed exactly \cite{BBDG} (see also section~\ref{sec_persistence}):}
\begin{eqnarray}\label{theta_of_beta}
&&\theta_+ = \frac{1}{2} - \frac{1}{\pi \alpha}\,\rm{arctan}\left(\beta \tan\left(\frac{\pi\alpha}{2}\right)\right) \;, \; \alpha \neq 1 \; \\
&&\theta_- = 1 - \theta_+ =  \frac{1}{2} + \frac{1}{\pi \alpha}\,\rm{arctan}\left(\beta \tan\left(\frac{\pi\alpha}{2}\right)\right) \;, \; \alpha \neq 1 \;.
\end{eqnarray} 

\begin{figure}[h]
\centering
\includegraphics[width=0.8 \columnwidth]{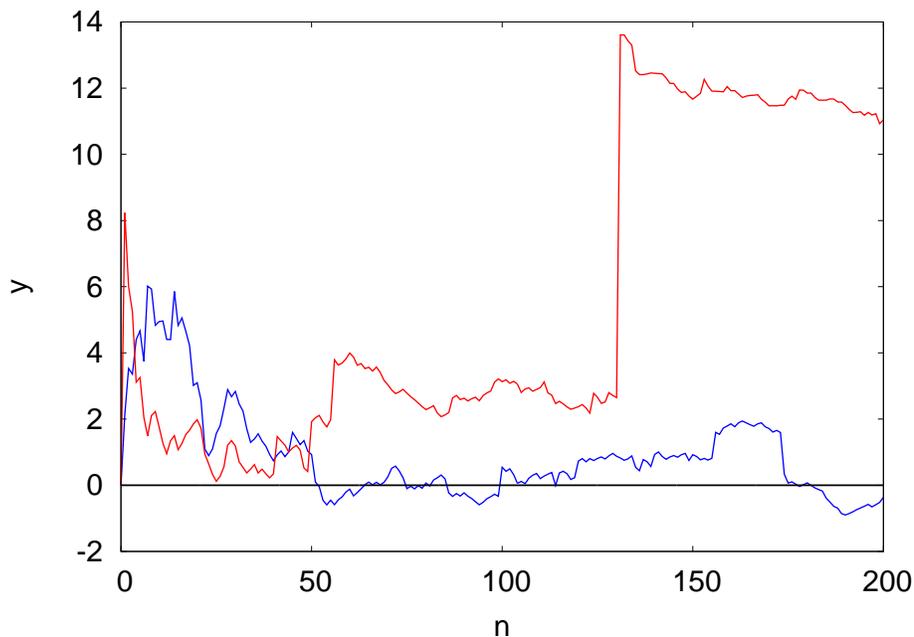}
\caption{Two one-dimensional L\'evy walkers after $n$ steps. In presence of an absorbing boundary condition in the negative half-line (under the black line), the blue walker has been absorbed and the other one (red) is still alive.}
\label{f:0c}
\end{figure}

{Here} we focus on the PDF of the rescaled variable $y$ (\ref{scaling2}) in the case where the walker is confined on the semi-axis $[0,+\infty)$  (Fig.~\ref{f:0c}), namely $R_+(y)$.
Far from the boundary this PDF,  $R_+(y)$,  displays the same
algebraic decay as the original jump distribution $\phi(\eta)$  (i.e.  $\propto y^{-1-\alpha}$)~\cite{Zumofen_Klafter},  but with a different amplitude  $c_+$ instead of $c$. Here we compute the exact value of the amplitude $c_+$ and show that it is related to the corresponding persistence exponent $\theta_+$ given in (\ref{theta_of_beta}) (see section~\ref{sec_tail_amp}):
\begin{eqnarray}\label{conjecture1D_c}
 R_+(y) \sim \frac{c_+}{y^{1+\alpha}}  \quad, \quad     c_+ = \frac{c}{1-\theta_+}\;,
\end{eqnarray}
This result is in agreement with the previous prediction $c_+=2 \, c$ valid only for symmetric L\'evy flights (where $\beta=0$ and $\theta_+=1/2$). In this case, this result was first obtained in  \cite{Rosso_Schehr} using a perturbative expansion around $\alpha=2$ \cite{zoia_rosso},   and confirmed by an exact calculation valid for any $\alpha$ in \cite{WMS12}.

This last result (\ref{conjecture1D_c}) can be generalized to more complex situations in a $d-$dimensional space where the walker is constrained to stay in a semi-bounded domain $\mathcal{D}$ (for instance a wedge in 2-$d$ or a cone in 3-$d$) and is absorbed if it jumps outside. In this case the survival probability has also an algebraic decay with a persistence exponent $\theta_\mathcal{D}$. Far from the boundaries the PDF  of the rescaled variable $\vec{y}$, $R_{d,\mathcal{D}}$, displays the same algebraic decay as the PDF $R_d$, in absence of boundaries. In this case, we show that the amplitudes of the decay are related via the persistence exponent (see section~\ref{sec_2D}):
\begin{eqnarray}\label{conjecture_nD}
\frac{R_{d,\mathcal{D}}(\vec{y})}{R_d(\vec{y})} \underset{{\rm d}(\vec y,\partial {\mathcal D}) \to \infty}{\longrightarrow} \frac{1}{1-\theta_\mathcal{D}}\;, 
\end{eqnarray}
where ${\rm d}(\vec y,\partial {\mathcal D})$ denotes the distance between the point located at $\vec y$ and the boundary of $\mathcal{D}$.
This result is based on a heuristic argument valid for fat-tail jump distributions and  is confirmed by numerical simulations in dimensions $d=1$ and $d=2$.

The paper is organized as follows. In section 2 we present our analytical results for asymmetric L\'evy flights on a one-dimensional half-line: {we first give a detailed derivation of the persistence exponents in section 2.1} and then we compute the tail of the constrained propagator $R_+(y)$ in section 2.2. In section 3, we confront our exact results in one-dimension to thorough numerical simulations and in section 4 we test our generalization for the tail of the propagator (\ref{conjecture_nD})
to the case of a L\'evy walker in a 2-$d$ wedge before we conclude in section 5.  

\section{Analytical results in one dimension}\label{section:theoretical}

\subsection{Persistence exponent}\label{sec_persistence}

\begin{figure}
\includegraphics[width= 0.52\columnwidth]{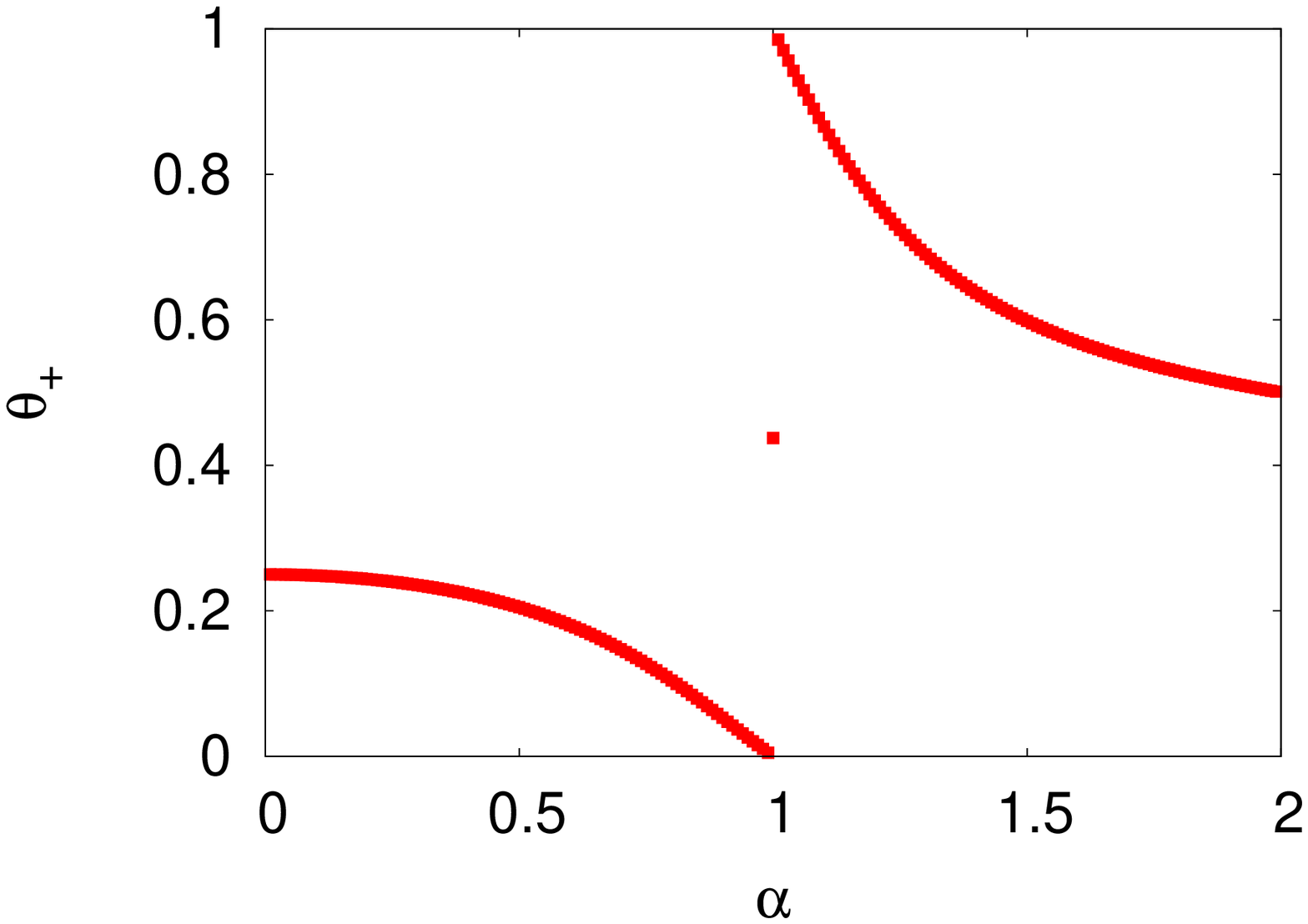}
\includegraphics[width= 0.52\columnwidth]{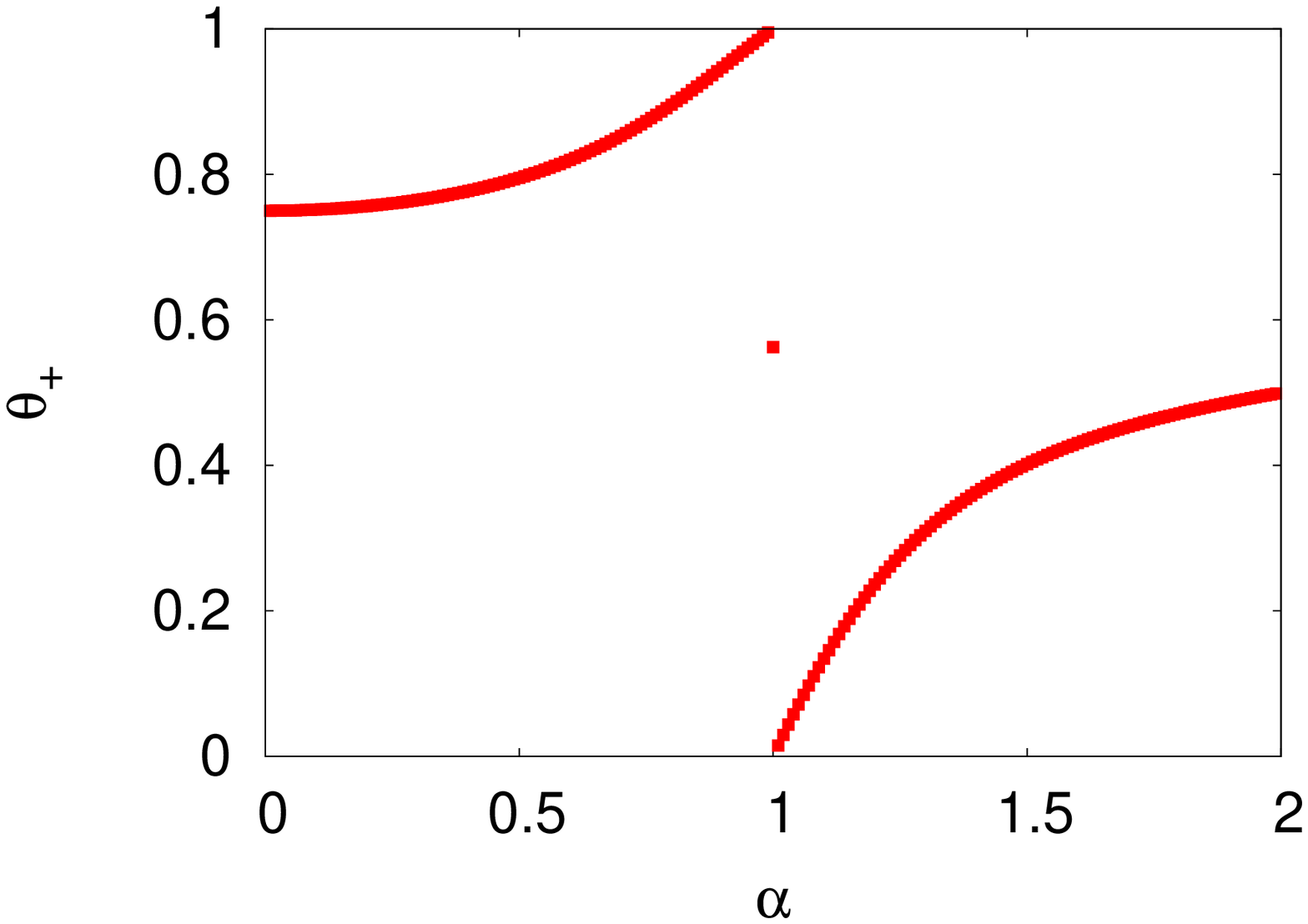}
\caption{{\bf Left:} Plot of $\theta_+(\alpha,\beta = 1/2)$ given in Eq. (\ref{theta_of_beta}). {\bf Right:} Plot of $\theta_+(\alpha,\beta = -1/2)$ given in Eq. (\ref{theta_of_beta}). For $\alpha = 1$, the values have been evaluated numerically from (\ref{relation_theta_rho}) and (\ref{def1b}).}\label{Fig:theta}
\end{figure}

We are interested in computing the survival probabilities $q_+(n)$ and $q_-(n)$ defined by Eqs. (\ref{q+}) and (\ref{q-}). The expression for the persistence exponents $\theta_{+}$ and $\theta_-$ was obtained in Ref. \cite{BBDG} in the different context of generalized persistence for spin models. We found it useful to give the details of the derivation of these results directly in terms of the survival
probability of random walks on the positive half-line. These survival probabilities $q_+(n)$ and $q_-(n)$ can be computed using the (generalized) Sparre Andersen theorem \cite{SA54} which yields explicit expressions for their generating functions, $\tilde q_{\pm}(s)$  as
\begin{eqnarray}\label{th_SA}
&&\tilde q_{+}(s) = \sum_{n=0}^\infty q_{+}(n) s^n = \exp{\left[ \sum_{n=1}^\infty \frac{p_n^+}{n} s^n \right]} \;, \; p_n^+ = {\rm Prob.}\,[x_n \geq 0] \;, \\
&&\tilde q_{-}(s) = \sum_{n=0}^\infty q_{-}(n) s^n = \exp{\left[ \sum_{n=1}^\infty \frac{p_n^-}{n} s^n \right]} \;, \; p_n^- = {\rm Prob.}\,[x_n \leq 0] \;. \nonumber
\end{eqnarray}
Note that in the symmetric case ($\beta = 0$), one has simply $p_n^+ = p_n^- = 1/2$ and, using $\sum_{n\geq 1} s^n/n = -\ln{(1-s)}$, this yields, for $\beta = 0$
\begin{eqnarray}\label{SA_std}
\tilde q_{+}(s) = \tilde q_-(s) = \frac{1}{\sqrt{1-s}} \Longrightarrow q_+(n) = q_-(n) = {2n \choose n} \frac{1}{2^{2n}} \underset{n\to\infty}{\sim} \frac{1}{\sqrt{\pi n}}\;, 
\end{eqnarray}
independently of the jump distribution. In the asymmetric case, $\beta \neq 0$, the situation is slightly more complicated and we focus now on $q_+(n)$. Its large $n$ behavior can be obtained by analysing the behavior of its  generating function when $s \to 1$. In the right hand side  of Eq. (\ref{th_SA}), the series in the argument of the exponential is dominated, when $s \to 1$, by the large $n$ terms. In this regime, one expects the scaling form in Eq.~(\ref{scaling2})  such that $p_n^+ \to \int_0^\infty R(y) dy$ for $n \to \infty$ implying that
\begin{eqnarray}\label{asympt_plus}
\sum_{n=1}^\infty \frac{p_n^+}{n} s^n \underset{s \to 1}{\sim} -\rho \ln{(1-s)} \;, \; \rho = \int_0^\infty R(y) dy \;,
\end{eqnarray} 
independently of the jump distribution $\phi(\eta)$ [with tails as in (\ref{tail_jump})]. Therefore from the Sparre Andersen theorem (\ref{th_SA}) and the above asymptotic result (\ref{asympt_plus}) one gets that $\tilde q_+(s) \sim (1-s)^{-\rho}$ and, from standard Tauberian theorem, 
\begin{eqnarray}\label{q_large_n}
q_+(n) \sim \frac{1}{\Gamma(\rho)} n^{-\theta_+} \;, \; \theta_+ = 1-\rho \;.
\end{eqnarray}
One can show similarly that $q_-(n) \sim n^{-\theta_- }/\Gamma(1-\rho)$ with $\theta_- =\rho$.
 Finally, using the expression of the characteristic function of $R(y)$  given in Eq. (\ref{def1b}) it is possible to compute explicitly $\rho$ (which is sometimes known under the name of the Zolotarev integrand) \cite{zolotarev_book}
\begin{eqnarray}\label{zolotarev_int}
\rho = \int_0^\infty R(y) dy = \frac{1}{2} + \frac{1}{\pi \alpha}  \rm{arctan}\left[\beta \tan\left(\frac{\pi\alpha}{2}\right)\right]\,,\; \alpha \neq 1 \;.
\end{eqnarray}
which, together with Eq. (\ref{q_large_n}), yields the expression for $\theta_+$ 
\begin{eqnarray}\label{relation_theta_rho}
\theta_+ = 1- \rho = \int_{-\infty}^0 R(y) \, dy \;, 
\end{eqnarray}
given in Eq. (\ref{theta_of_beta}) for $\alpha \neq 1$. For $\alpha = 1$, the exponent $\theta$ can be evaluated numerically from (\ref{relation_theta_rho}) and (\ref{def1b}). In Fig. \ref{Fig:theta} we show a plot of the exact formula of $\theta_+(\alpha, \beta)$ for $\beta = 1/2$ (left) and $\beta = -1/2$ (right), given in Eq. (\ref{theta_of_beta}) for $\alpha \neq 1$. In both cases, we observe that $\theta_+$ exhibits a discontinuity at $\alpha = 1$. This discontinuity can be traced back to the discontinuous behavior of the L\'evy stable distribution itself in (\ref{def1b}), for $\beta \neq 0$ as $\alpha$ crosses the value $\alpha = 1$. Note that a similar discontinuous behavior, for $\alpha =1$, was also observed in the numerical estimate of the mean first passage of skewed L\'evy flights in bounded domains \cite{hanggi}. 

Note that if the jump distribution 
$\phi(\eta)$ is itself a stable law, $\phi(\eta) = R(\eta)$, then one has $p^+_n = \rho$ exactly for all $n$ (not only in the asymptotic large $n$ limit), such that $\tilde q_+(s) = (1-s)^{-\rho}$ and in this case $q_+(n)$ can be computed exactly for all $n$ (see also \cite{SiAu12})
\begin{eqnarray}\label{exact_q}
q_+(n) = \frac{\Gamma(n+\rho)}{n! \, \Gamma(\rho)} \;, \; {\rm for} \; \phi(\eta) = R(\eta) \;.
\end{eqnarray}
Of course, for $\rho = 1/2$, one recovers the standard result of Sparre Andersen~\cite{SA53} (\ref{SA_std}). 

\subsection{Tail of the propagator with an absorbing boundary at the origin}\label{sec_tail_amp}


We are now interested in the asymptotic behavior of the distribution of the rescaled position $y = x_n/n^{1/\alpha}$ (\ref{scaling2}) of the walker given that it has survived up to time $n$, namely $R_+(y)$. For this purpose it is useful to introduce $F(M,n)$, the probability to find a free particle in $x> M$ after $n$ steps, and $F_+(M,n)$, the probability to find the constrained particle in $x> M$ after $n$ steps, where $M$ is a positive number. In terms of the scaling variable $x_n/n^{1/\alpha}$~(\ref{scaling2}), these probabilities can be expressed using $R$ and $R_+$ as
\begin{align}
F(M,n) &= \int_M^\infty \frac{{\rm d}x}{n^{1/\alpha}} \,R\left(\frac{x}{n^{1/\alpha}}\right) = \int_{M/n^{1/\alpha}}^\infty {\rm d}y \,R\left(y\right)\;,\label{def_F}\\
\nonumber F_+(M,n) &= \int_M^\infty \frac{{\rm d}x}{n^{1/\alpha}} \,R_+\left(\frac{x}{n^{1/\alpha}}\right) = \int_{{M}/{n^{1/\alpha}}}^\infty {\rm d}y \,R_+\left(y\right)\;.
\end{align}
We are interested in the behavior of $F(M,n)$ and $F_+(M,n)$ when $M/n^{1/\alpha} \gg 1$ so that one can use the asymptotic behaviors of $R(y)$ and $R_+(y)$ to evaluate the integrals in (\ref{def_F}): $R(y)\sim c/y^{1+\alpha}$ and $R_+(y)\sim c_+/y^{\alpha+1}$ when $y \to + \infty$. 
We obtain the asymptotic behavior of $F(M,n)$ and $F_+(M,n)$ in the limit of large $M / n^{1/\alpha}$ ($M \gg n^{1/\alpha} \gg 1$),
\begin{eqnarray}
F(M,n) \underset{M \gg n^{1/\alpha} \gg 1}{\sim} \frac{n}{\alpha} \frac{c}{M^\alpha} \qquad {\rm and} \qquad F_+(M,n) \underset{M \gg n^{1/\alpha} \gg 1}{\sim} \frac{n}{\alpha} \frac{c_+}{M^\alpha} \;.
\end{eqnarray}
Therefore we get
\begin{eqnarray}\label{c+_c}
\frac{c_+}{c} = \lim_{n \to \infty} \lim_{M \to \infty}\frac{F_+(M,n)}{F(M,n)}\;.
\end{eqnarray}
To compute the right hand side of this equation, we write formally $F_+(M,n)$ as
\begin{eqnarray}\label{formal_F+}
F_+(M,n) &=& {\rm Prob.}\bigl[x(n)>M \vertt \forall n'\in[0,n],\,x(n')>0\bigr]\;,
\end{eqnarray}
where we denote by ${\rm Prob.}(A|B)$ the probability of $A$ given $B$. We then develop this formula (\ref{formal_F+})
using Bayes' formula \footnote{${\rm Prob.}(A|B) {\rm Prob.}(B) = {\rm Prob.}(B|A) {\rm Prob.}(A) = {\rm Prob.}(A \cap B)$ which implies ${\rm Prob.}(A|B)  = {\rm Prob.}(B|A) {\rm Prob.}(A)/{\rm Prob.}(B)$}
\begin{align}
F_+(M,n) &=& \frac{{\rm Prob.}\bigl[x(n)>M\bigr]}{{\rm Prob.}\bigl[\forall n'\in[0,n],\,x(n')>0\bigr]}\,{\rm Prob.}\bigl[\forall n'\in[0,n],\,x(n')>0 \vertt x(n)>M\bigr]\;.\label{F/S}
\end{align}
Here we recognize the probability, in the numerator, $F(M,n) = {\rm Prob.}\bigl[x(n)>M\bigr]$ and, in the denominator, the survival probability $q_+(n)= {\rm Prob.}\bigl[\forall n'\in[0,n],\,x(n')>0\bigr]$.
To evaluate ${\rm Prob.}[\forall n'\in[0,n],\,x(n')>0 \vertt x(n)>M]$ in the limit of large $M$, we assume that the trajectories such that $x(n) > M$ are characterized by a single jump larger than $M$ which happens at a step $n_1$ which may occur at any time in the interval $[0,n]$, hence $\eta(n_1) > M$. Thus, after this big jump the particle stays above $0$ with a probability $1$ as it is already far away from the origin. This argument, namely the fact that the trajectory is dominated by a single large jump, holds only for jump distributions with heavy tails ($\alpha< 2$), (hence it does not hold for standard random walks which converge to Brownian motion). Within this hypothesis we obtain [using ${\rm Prob.}(A|B) = {\rm Prob.}(A\cap B)/{\rm Prob.}(B)$]
\begin{align}\label{hypothese}
{\rm Prob.}[\forall n'\in[0,n],\,x(n')>0 \vertt x(n)>M] \underset{M \to \infty}{\sim} \frac{\displaystyle\sum_{n_1=0}^{n} q_+(n_1)\;{\rm Prob.} [\eta(n_1)>M]}{\displaystyle\sum_{n_1=0}^n {\rm Prob.} [\eta(n_1)>M]}\;.
\end{align}
As the jumps variables are i.i.d., ${\rm Prob.} [\eta(n_1)>M]$ is independent of $n_1$. Therefore we find:
\begin{eqnarray}\label{conjecture_th}
\lim_{n \to \infty} \lim_{M \to \infty} \frac{F_+(M,n)}{F(M,n)} = \lim_{n \to \infty} \frac{\displaystyle\sum_{n_1=0}^{n} q_+(n_1)}{n\,q_+(n)}\;.
\end{eqnarray}

\noindent Then, replacing $q_+(n_1)$ by its expression in Eq.~(\ref{q_large_n}), we get,
\begin{eqnarray}
\lim_{n \to \infty} \lim_{M \to \infty} \frac{F_+(M,n)}{F(M,n)} = \frac{1}{1-\theta_+} \;,\; {\rm for}\; \theta_+ < 1,
\end{eqnarray}
which finally leads, with Eq.~(\ref{c+_c}), to the general result (\ref{conjecture1D_c}) in one dimension.\\

\section{Numerical simulations in one dimension}

To test our predictions for the persistence exponent and the tail of the constrained propagator, we have simulated numerically the random walk defined by Eq. (\ref{def_rw}). In our simulations, we chose for the jump distribution $\phi(\eta)$ the Pareto distribution (see the left panel of Fig.~\ref{f:Pareto_f:1}) -- which is a fat tailed distribution, easier to handle numerically than a stable law. It is defined for a positive $\alpha$ by (see also the left panel of Fig. \ref{f:Pareto_f:1}):
\begin{eqnarray}\label{def_pareto}
\phi(\eta)= 
	\begin{cases}
		\displaystyle\frac{c}{\eta^{\,\alpha+1}} &{\rm for} \; \eta > b_+ \;,\vspace{1mm}\\
		\displaystyle\frac{c/\gamma}{|\eta|^{\alpha+1}} & {\rm for} \; \eta < -b_-\;, \;{\rm where}\; \gamma = \frac{1+\beta}{1-\beta},\vspace{1mm}\\
		0 &{\rm otherwise}\;,   
	\end{cases}
\end{eqnarray}
This distribution must be normalized and have a mean equal to $\mu = 0$. These two conditions give us $b_-$ and $b_+$as:
\begin{eqnarray}\label{a_b_pareto}
({b_-})^\alpha = \frac{c\,(1+\gamma^{\frac{1}{1-\alpha}})}{\alpha \,\gamma}\qquad {\rm and} \qquad
({b_+})^\alpha = \frac{c\,(1+\gamma^{1-\alpha})}{\alpha} \;.
\end{eqnarray}
\noindent  For $\alpha$ in $(0,2)$ the process converges to a skewed L\'evy stable process with stability index $\alpha$, skewness parameter $\beta$ and  a scale parameter $c$.

To generate random jump variables $\eta$ distributed according to a Pareto law we can use the direct sampling method~\cite{Krauth}. 
In  practice, at each step, the walker makes a positive jump with a probability $\pi_+ = \int_{b_+}^{+\infty} {\rm d}\eta\, \phi(\eta)$, and a negative jump with a probability $1-\pi_+$. The amplitude of this jump is then given by~\cite{Krauth}
\begin{eqnarray}\label{pareto_bis}
\eta= 
\begin{cases}
{\rm rand}(0, {(b_+)}^{-\alpha})^{-\frac{1}{\alpha}}\;,&\qquad {\rm with}\;{\rm probability} \; \pi_+\;,\vspace{3mm}\\
- {\rm rand}(0, {(b_-)}^{-\alpha})^{-\frac{1}{\alpha}}\;,&\qquad {\rm with}\;{\rm probability} \; 1-\pi_+\;,
\end{cases}
\end{eqnarray}
where rand$(x,y)$ is a random number drawn randomly from a uniform distribution in the interval $(x,y)$. We first present our results for the persistence exponent and then for the tail of the constrained propagator.

\subsection{Survival probability and persistence exponent}

\begin{figure}[h]
\begin{center}
\includegraphics[width=0.49 \columnwidth]{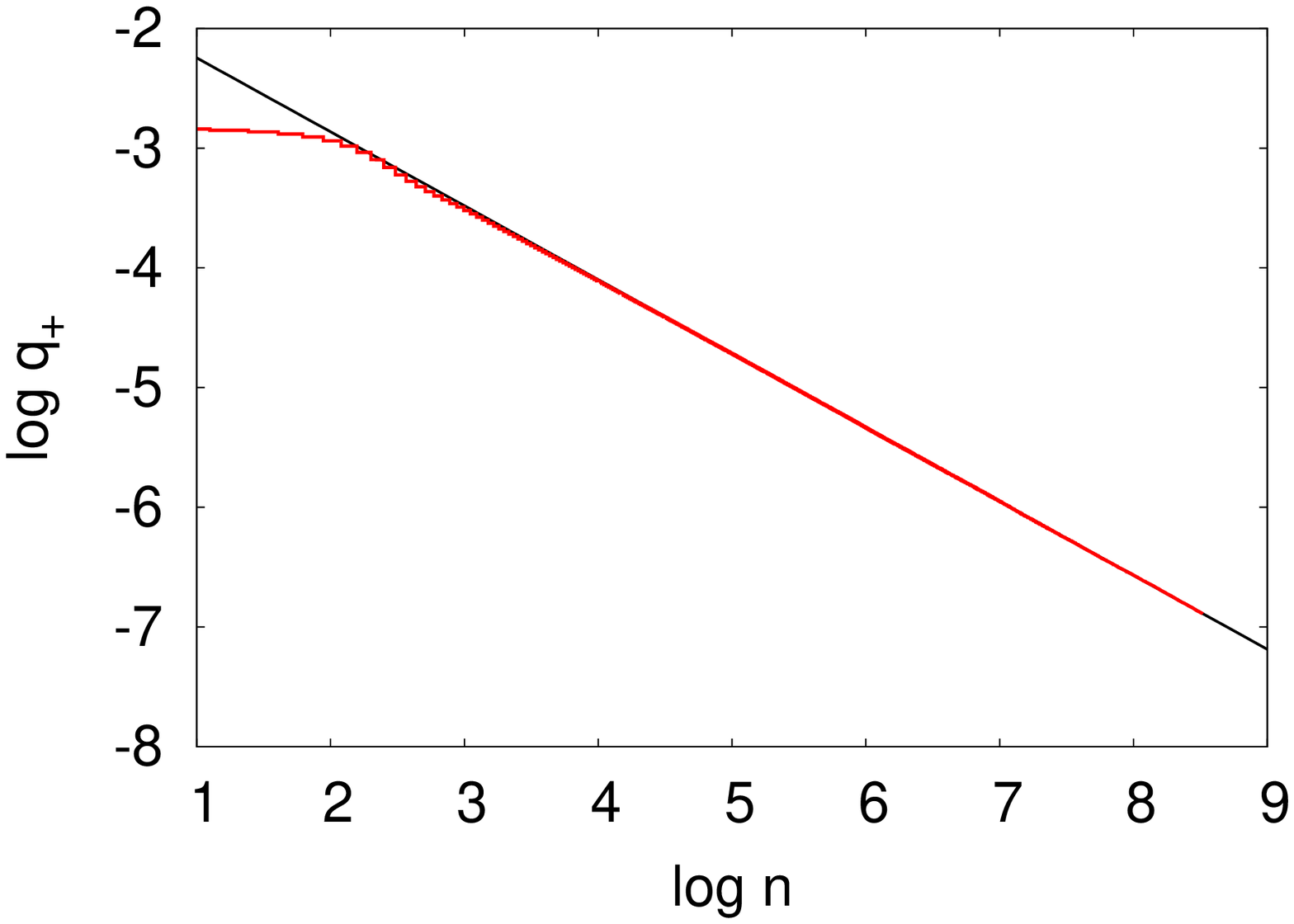} 
\includegraphics[width=0.49 \columnwidth]{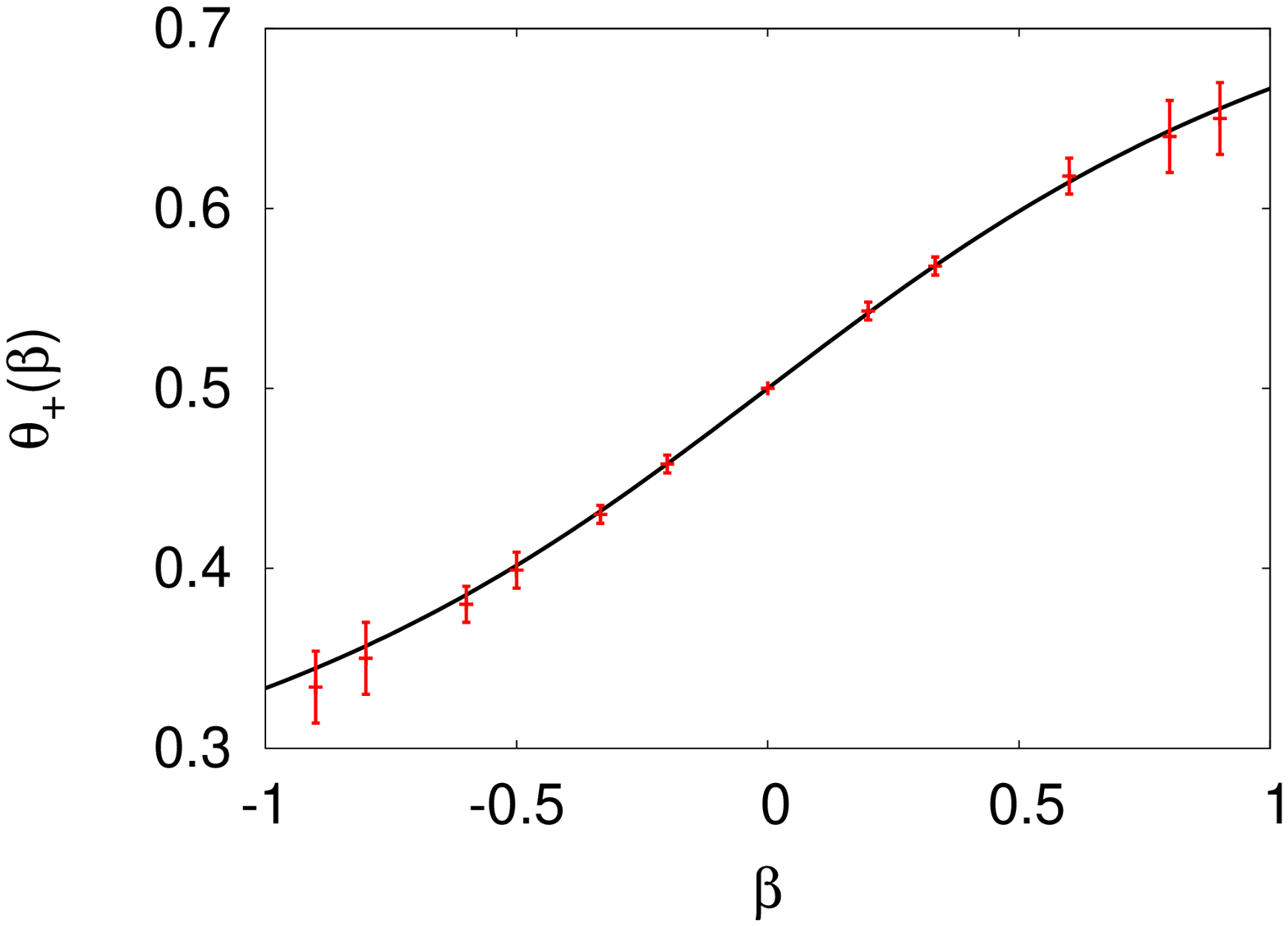}
\caption{{\bf Left}: Survival probability of an asymmetric L\'evy random walk (with $\alpha = 3/2$, $\gamma=4$, $c = 3$ and $n=5000$) constrained on the positive half axis. This graphic is performed using $10^7$ samples. A fit of the tail (black) gives us $\theta_+ = 0.62 \pm 0.01$. {\bf Right}: Plot of the persistence exponent $\theta_+$ as a function of $\beta = (\gamma-1)/(\gamma+1)$, from the data given in Table \ref{table1} (here $\alpha = 3/2$). The red symbols are the numerical estimates of $\theta_+$ extracted from the algebraic decay of $q_+(n)$ while the solid line is our exact analytical result 
$\theta_+(\beta) = \frac{1}{2}-\frac{1}{\pi\,\alpha}\,\arctan(\beta\tan(\frac{\pi\,\alpha}{2}))$ in Eq. (\ref{theta_of_beta}).}
\label{f:logS_logR+}
\end{center}
\end{figure}

To compute the survival or persistence probability $q_+(n)$, defined in Eq. (\ref{q+}), we generate a large number of independent L\'evy walkers, 
evolving via Eqs. (\ref{def_rw}) and (\ref{pareto_bis}), and compute the fraction of walkers which remained on the positive axis until step $n$. In the left panel of Fig.~\ref{f:logS_logR+} we show a plot of $q^+_n$ as function of $n$ in a log-log scale for $\alpha = 3/2$ and $\gamma=4$ (corresponding to $\beta = 3/5$). The straight line observed on this log-log plot is in full agreement with the expected algebraic decay, $q^+_n \propto n^{-\theta_+}$. From these data one can extract a reliable numerical estimate of the exponent $\theta_+$. 


We have then measured the persistence probability for different values of $\gamma$ (asymmetry of the distribution) and for $\alpha = 3/2$, which allowed us to extract the persistence exponent as $\gamma$ (or $\beta = (\gamma-1)/(\gamma+1)$) is varied (see Table \ref{table1}).
\begin{table}
\centering
\begin{tabular*}{0.6\textwidth}{@{\extracolsep{\fill}}|c|c||c|c|}
  \hline
  $\gamma$ & $\beta$ & numerical $\theta_+$ & exact $\theta_+$\\
  \hline
  \hline
  $1/19$ & $-0.9$ & $0.33\pm0.02$ & $0.344 \ldots$\\
  \hline
  $1/9$ & $-0.8$ & $0.35\pm0.02$ & $0.357 \ldots$\\
  \hline
  $1/4$ & $-0.6$ & $0.38\pm0.01$ & $0.385 \ldots$\\
  \hline
  $1/3$ & $-0.5$ & $0.40\pm0.01$ & $0.402 \ldots$\\
  \hline
  $1/2$ & $-1/3$ & $0.430\pm0.005$ & $0.4317 \ldots$\\  
  \hline
  $2/3$ & $-0.2$ & $0.458\pm0.005$ & $0.4581 \ldots$\\
  \hline
  $1$ & $0$ & $0.5$ & $0.5$\\
  \hline
  $3/2$ & $0.2$ & $0.543\pm0.005$ & $0.5419 \ldots$\\
  \hline
  $2$ & $1/3$ & $0.568\pm0.005$ & $0.5683 \ldots$\\
  \hline
  $4$ & $0.6$ & $0.62\pm0.01$ & $0.615 \ldots$\\
  \hline
  $9$ & $0.8$ & $0.64\pm0.02$ & $0.643 \ldots$\\
  \hline
  $19$ & $0.9$ & $0.65\pm0.02$ & $0.656 \ldots$\\
  \hline
\end{tabular*}
\caption{Summary of our numerical estimates for $\theta_+$, extracted from the algebraic decay of the persistence probability.}\label{table1}
\end{table}
In the right panel of Fig. \ref{f:logS_logR+}, we have plotted these numerical estimates of $\theta_+$ as a function of $\beta$ and compare it to our exact analytical results given by Eq.~(\ref{theta_of_beta}). As we can see, the agreement between both is very good. 
%

\subsection{Tail of the propagator}

We first check that our numerical procedure (\ref{def_rw}) and (\ref{pareto_bis}) yields back the correct free propagator $R(y)$ before we compute the constrained one, $R_+(y)$. 

\subsubsection{Free L\'evy walkers}

We construct a large number of independent L\'evy walks evolving via Eqs. (\ref{def_rw}) and (\ref{pareto_bis}). For each random walk we record the final position $x_n$ after $n$ steps, and compute $R(y)$, the histogram of the corresponding rescaled variable $y = x_n/n^{1/\alpha}$. According to the Central Limit Theorem, for a large number of steps $n$, the probability distribution of $y$ is expected to converge to the stable distribution $R(y)$, with the asymptotic expansion:
\begin{eqnarray}\label{convergence_R}
R(y) \rightarrow
\begin{cases}
	\displaystyle\frac{c}{y^{1+\alpha}}+\mathcal{O}\left(\frac{1}{y^{1+2\alpha}}\right)\;,\qquad {\rm if}\; y > 0 \;,
	\vspace{3mm}\\
	\displaystyle\frac{c/\gamma}{|y|^{1+\alpha}}+\mathcal{O}\left(\frac{1}{|y|^{1+2\alpha}}\right)\;,\qquad {\rm if}\; y < 0 \;.
\end{cases}
\end{eqnarray}

\begin{figure}[h]
\includegraphics[width=0.50 \columnwidth]{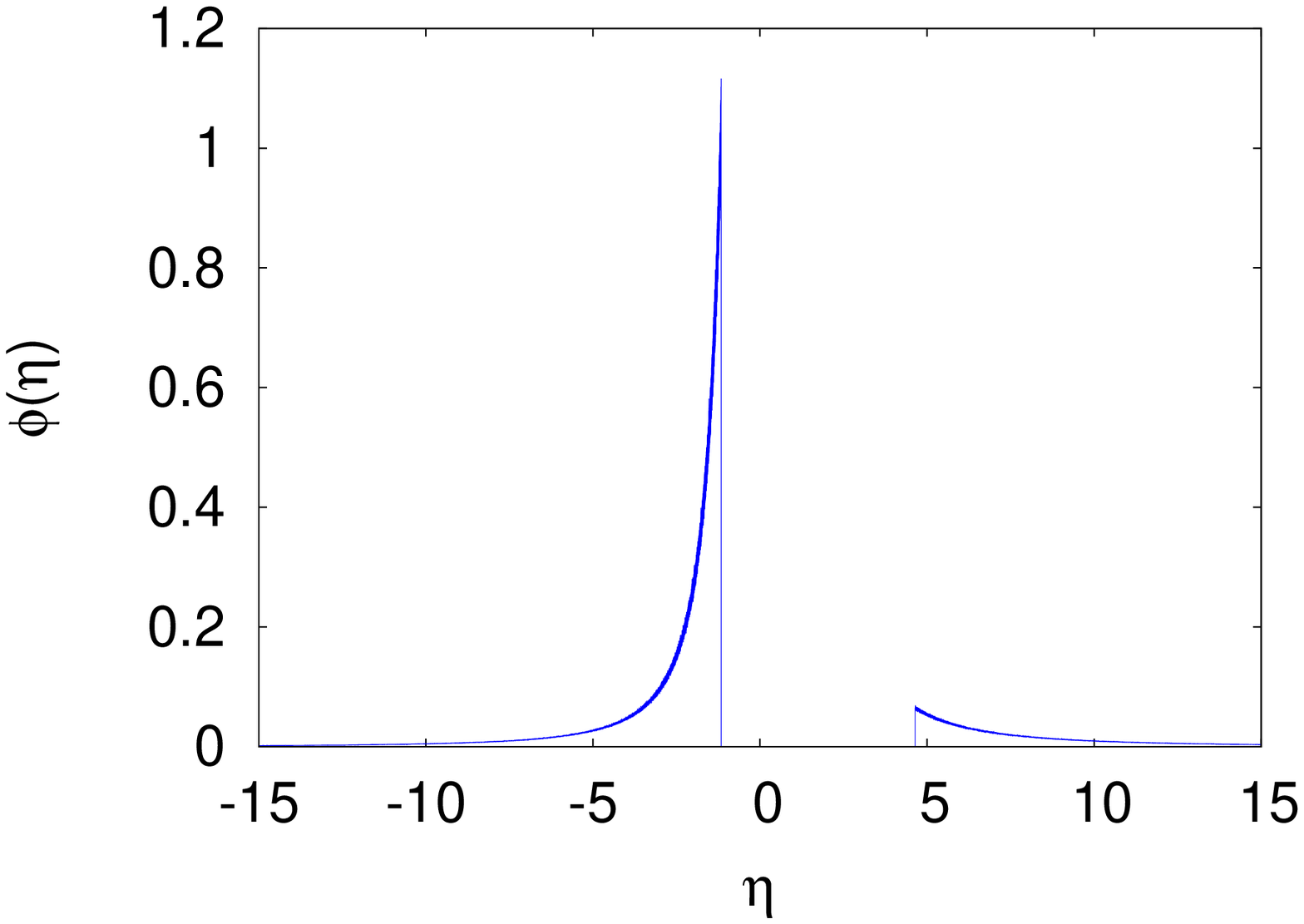} 
\includegraphics[width=0.50 \columnwidth]{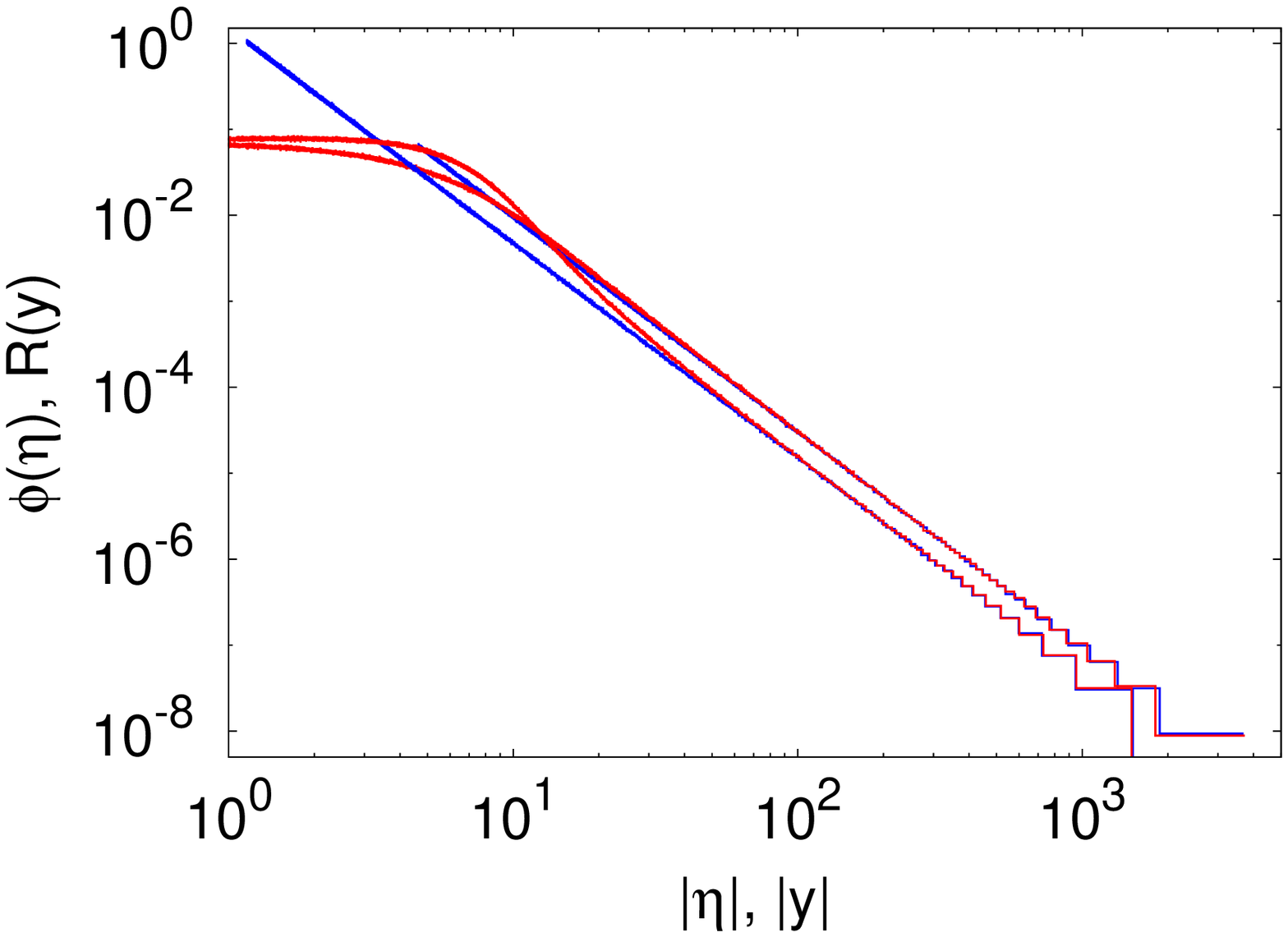}
\caption{{\bf Left}: Asymmetric Pareto distribution for $\alpha = 3/2$, $\gamma = 2$ and $c = 3$. Histograms are performed using $10^8$ samples. {\bf Right}: Comparison between Pareto (blue) and $R(y)$ (red) for a L\'evy flight of $n = 1000$ steps with the same parameters as in the left panel. The right and the left tails of the two PDFs have the same algebraic behavior.}
\label{f:Pareto_f:1}
\end{figure}

\noindent Our simulations recover this expected result: in the right panel of Fig.~\ref{f:Pareto_f:1}, we show that the tail of $R(y)$ coincides with the tail of $\phi(\eta)$ when, respectively, $y$ and $\eta$ are large.\\

\subsubsection{Constrained L\'evy walkers}

\begin{figure}[h]
\begin{center}
\includegraphics[width=0.55 \columnwidth]{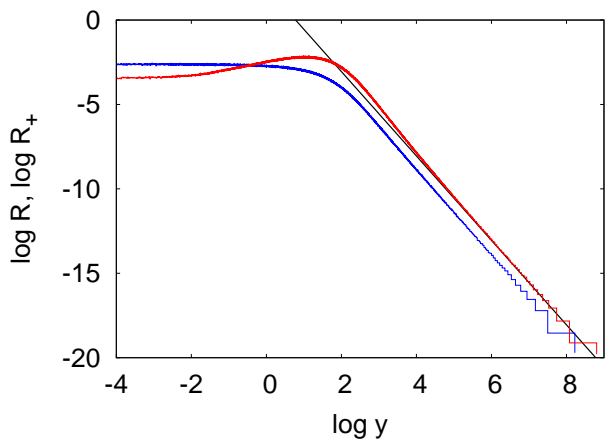}
\caption{PDFs of the rescaled variable $y$ for L\'evy flights with $\alpha = 3/2$, $\gamma = 2$, $c = 3$ and $n = 1000$ steps ($10^8$ samples) in presence, $R_+(y)$ (red), and in the absence, $R(y)$ (blue), of an absorbing boundary on the negative half-axis. $R_+(y)$ is compared to its expected asymptotic behavior $R_+^{\rm exp}(y)$ (black) given in Eq. (\ref{rexp}).}
\label{f:2}
\end{center}
\end{figure}

We now consider a one-dimensional random walk constrained to stay positive (Fig.~\ref{f:0c}). If the particle has survived on the positive semi-axis up to step $n$, we record its final position $x_n$. Then we construct $R_+(y)$, the histogram of the rescaled final positions ($y = x_n/n^{1/\alpha}$) from a large number of such constrained walks. In Fig.~\ref{f:2} we show a plot of both $R(y)$ and $R_+(y)$ (which is defined only for positive $y$) on a log-log scale. These two functions have the same asymptotic behavior but $R_+(y)$ is shifted from $R(y)$. This confirms that $R$ and $R_+$ both decay as $\propto y^{-\alpha-1}$ when $y$ becomes large (Ref.~\cite{Zumofen_Klafter}), but with different amplitudes ($c_+ \neq c$). 
%
%
%
We can now verify our main result in Eq.~(\ref{conjecture1D_c}), $c_+/c = 1/(1-\theta_+)$ as we know (exactly) the exponent~$\theta_+$.


In Fig. \ref{f:2} we compare the tail of $R_+$ to its expected tail $R_+^{\rm exp}$~(\ref{conjecture1D_c}),  for $\alpha=3/2$ and $\gamma = 2$:
\begin{eqnarray}\label{rexp}
\;\; R_+(y) \underset{y \to +\infty}{\sim} \frac{c_+}{y^{5/2}} \qquad \qquad R_+^{\rm exp}(y) \underset{y \to +\infty}{\sim} \frac{c}{1-\theta_+}\,\frac{1}{y^{5/2}}\;.
\end{eqnarray}
This expected tail fit very well $R_+(y)$ when $y$ becomes large, which is consistent with the relation~(\ref{conjecture1D_c}) for asymmetric cases in one dimension. A more precise comparison can be made from the evaluation of $c_+$ by fitting the algebraic tail of $R_+(y)$, which yields $c_+/c = 2.34 \pm 0.05$ while our exact result predicts $1/(1-\theta_+) \simeq 2.316\dots$ (taking the exact value of $\theta_+ = 0.5683\dots$). We have carried out simulations for different values of $\beta$ and extracted the amplitude $c_+$ of the tail. In Table \ref{table2} we compare these estimates of $c_+$ with the values of $1/(1-\theta_+)$. This comparison gives a good support to our heuristic argument leading to the relation in Eq. (\ref{conjecture1D_c}).  

\begin{table}
\centering
\begin{tabular*}{0.6\textwidth}{@{\extracolsep{\fill}}|c|c||c|c|}
  \hline
  $\gamma$ & $\beta$ & numerical $c_+/c$ & exact $c_+/c$\\
  \hline
  \hline
  $1/19$ & $-0.9$ & $1.52\pm0.05$ & $1.526 \ldots$\\
  \hline
  $1/9$ & $-0.8$ & $1.53\pm0.03$ & $1.555 \ldots$\\
  \hline
  $1/4$ & $-0.6$ & $1.61\pm0.03$ & $1.627 \ldots$\\
  \hline
  $1/2$ & $-1/3$ & $1.75\pm0.03$ & $1.760 \ldots$\\
  \hline
  $1$ & $0$ & $2$ & $2$\\
  \hline
  $2$ & $1/3$ & $2.34\pm0.05$ & $2.316 \ldots$\\
  \hline
  $4$ & $0.6$ & $2.6\pm0.1$ & $2.595 \ldots$\\
  \hline
  $9$ & $0.8$ & $3.0\pm0.5$ & $2.803 \ldots$\\
  \hline
  $19$ & $0.9$ & $3.3\pm0.5$ & $2.903 \ldots$\\
  \hline
\end{tabular*}
\caption{Summary of our numerical estimates for $c_+/c$, extracted from the algebraic tail of the constrained propagator $R_+(y)$ and compared to its expected value $\frac{c_+}{c} = \frac{1}{1-\theta_+}$ (taking the exact value of $\theta_+$ in Eq.~(\ref{theta_of_beta})).}\label{table2}
\end{table}

\section{Generalization to two-dimensional random walkers}\label{sec_2D}

The result in Eq. (\ref{conjecture1D_c}), valid for a one-dimensional L\'evy walker, can be generalized to $d$-dimensional L\'evy walkers constrained to stay within an open domain $\mathcal{D}$. Following the lines of reasoning presented in section \ref{section:theoretical}, we predict that far from the boundary the PDF $R_{d,\cal D}(\vec y)$ behaves like the PDF $R_d(\vec y)$ in absence of boundary with the universal ratio:
\begin{eqnarray}\label{conjecture_nD_section}
\frac{R_{d,{\mathcal{D}}}(\vec{y})}{R_d(\vec{y})} \underset{{\rm d}(\vec y,\partial {\mathcal D}) \to \infty}{\longrightarrow} \frac{1}{1-\theta_\mathcal{D}}\;, 
\end{eqnarray}
where ${\rm d}(\vec y,\partial {\mathcal D})$ denotes the distance between the point located at $\vec y$ and the boundary of $\mathcal{D}$. In Eq.~(\ref{conjecture_nD_section}), $\theta_{\cal D}$ is the persistence exponent defined via the survival probability $q_{\mathcal{D}}(n)$, i.e. the fraction of walkers which stay inside the domain ${\cal D}$ up to step $n$. Analogously to the one-dimensional case Eq.~(\ref{asympt_S}), when the number of jumps $n \to \infty$, $q_{\mathcal{D}}(n) \propto n^{-\theta_{\cal D}}$ (while there exists no exact result for $\theta_{\cal D}$).

\begin{figure}[h]
\begin{center}
\includegraphics[width=0.49 \columnwidth]{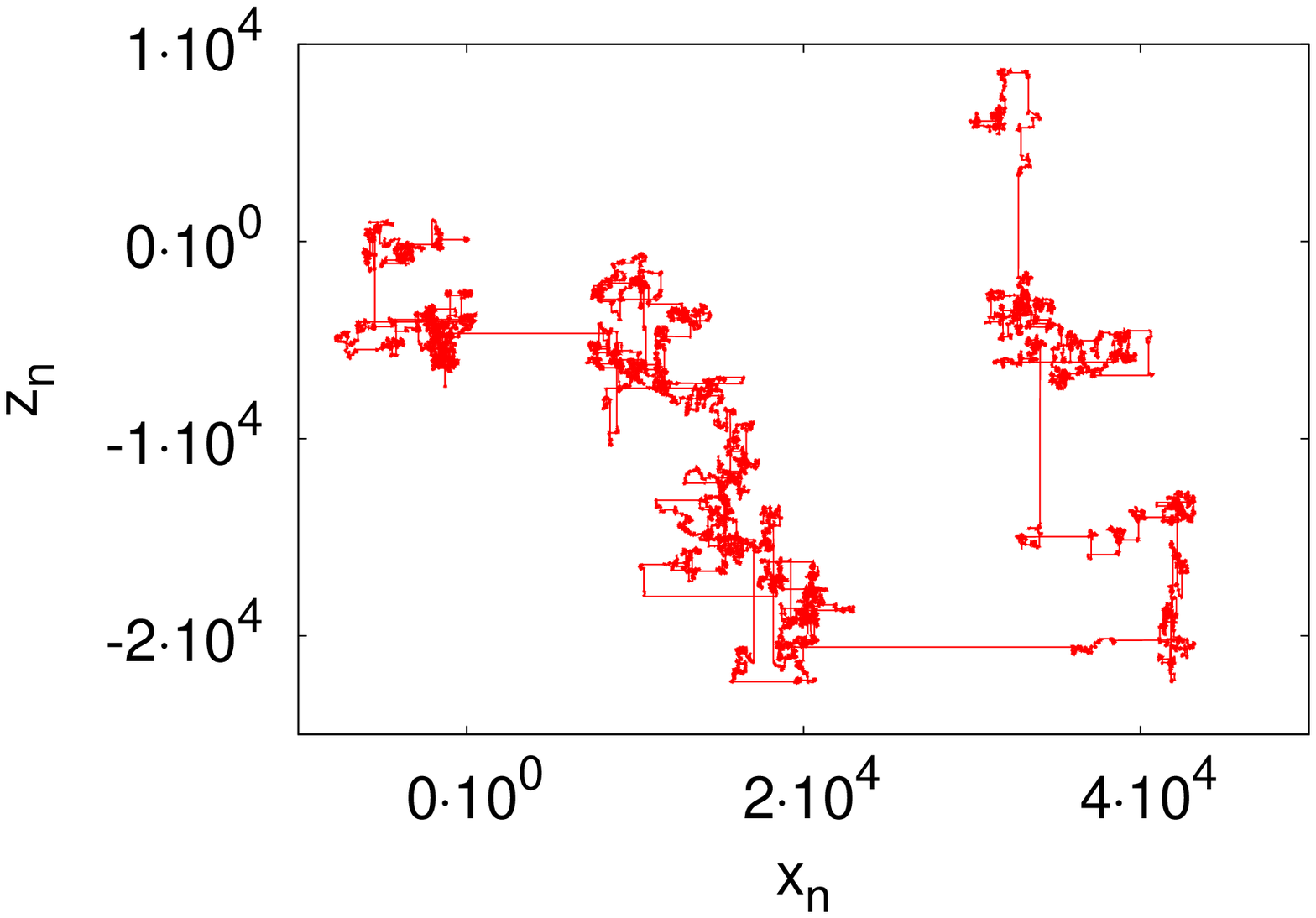}
\includegraphics[width=0.49 \columnwidth]{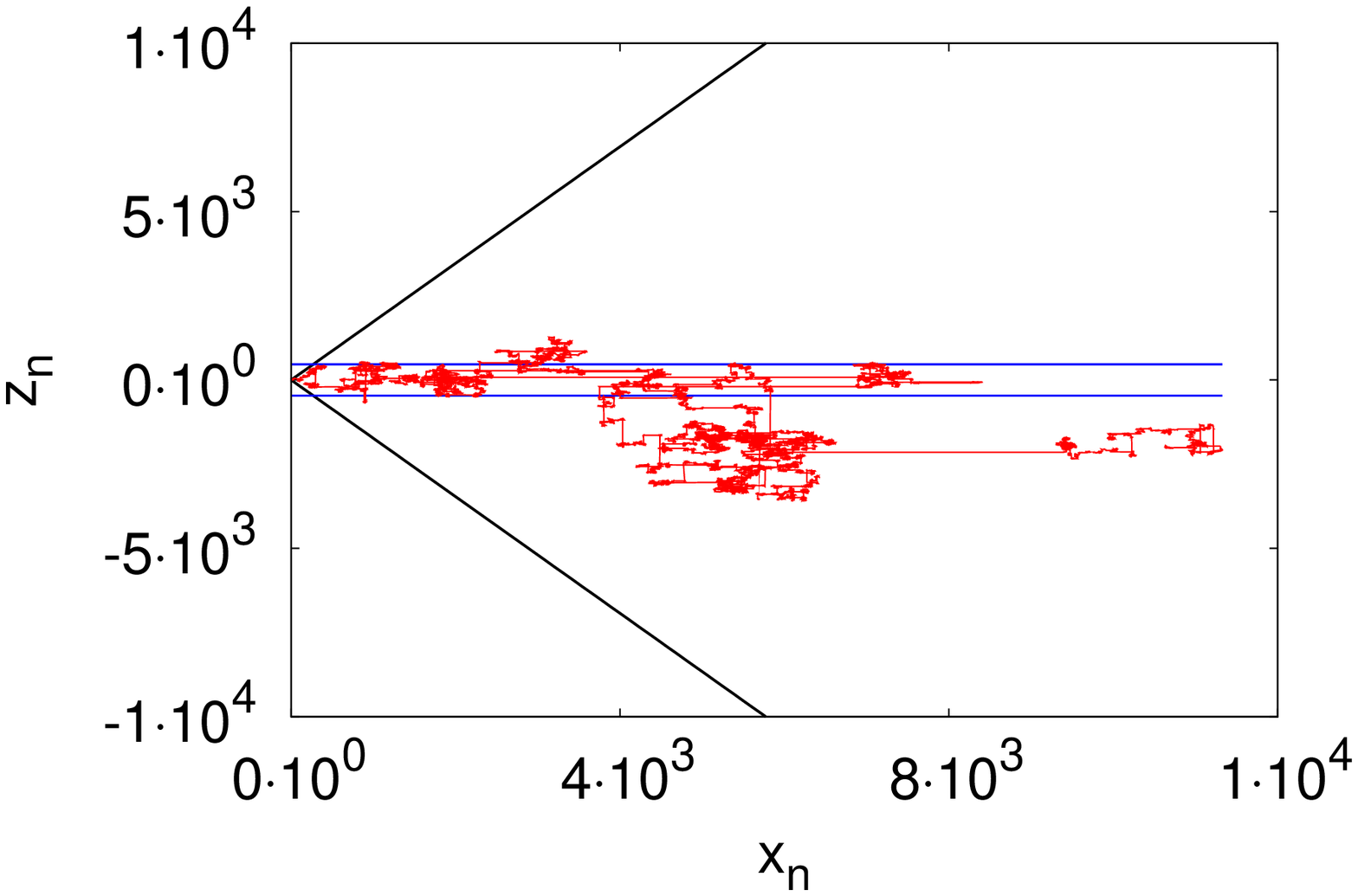}
\caption{$\alpha = 3/2$, $\gamma = 1$ and $c = 1$. {\bf Left}: Free L\'evy walker of $10^6$ steps in a two-dimensonal space, evolving according to (\ref{2D_adim}). {\bf Right}: A two-dimensional walk in the presence of the absorbing wedge (black). The blue lines delimit the stripe of width $2 \epsilon$ used to compute $R_d(u,0)$ and $R_{d,{\cal D}}(u,0)$ (\ref{numerical}).}
\label{f:0b:0d}
\end{center}
\end{figure}

\begin{figure}[h]
\begin{center}
\includegraphics[width=0.49 \columnwidth]{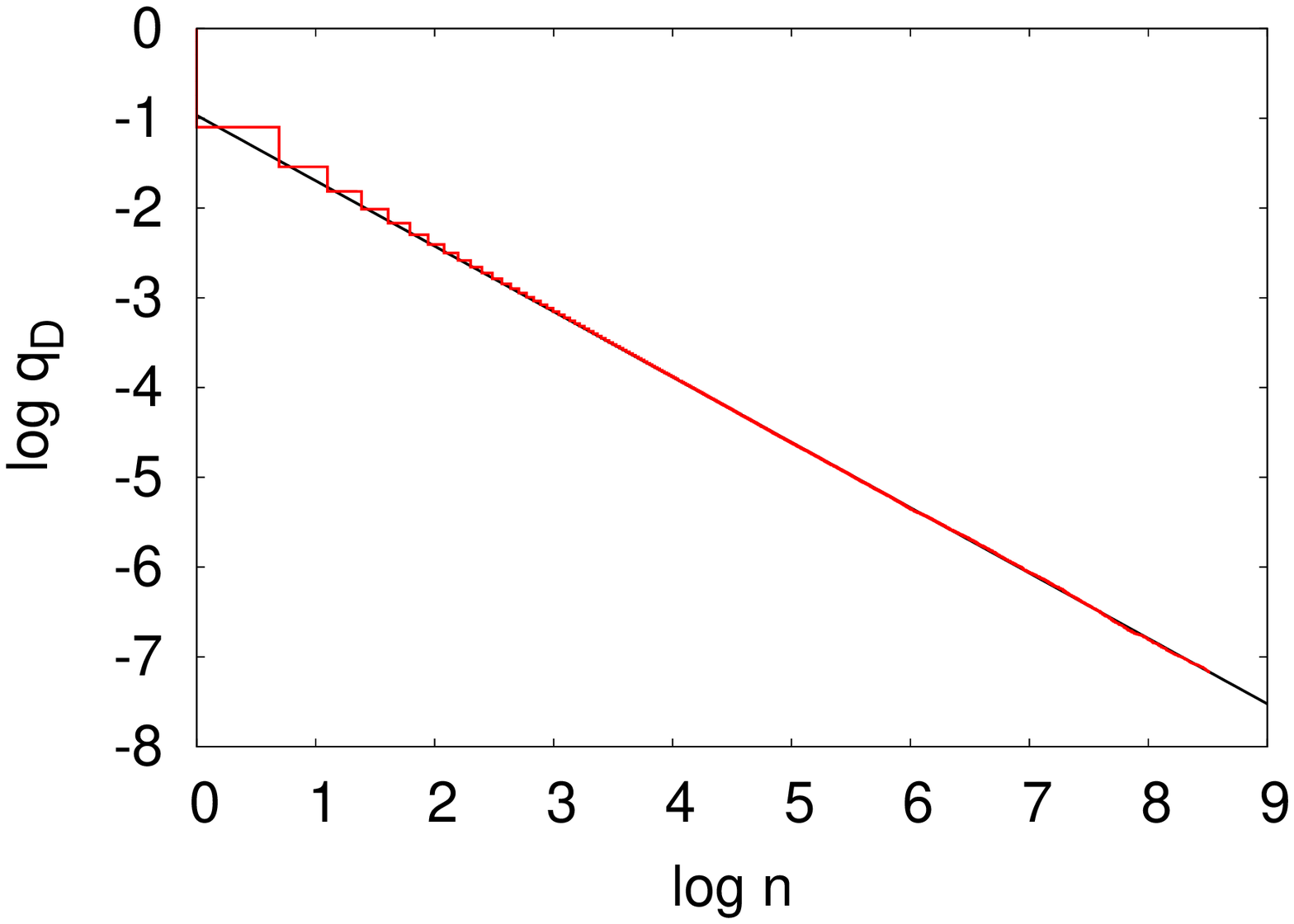}
\includegraphics[width=0.49 \columnwidth]{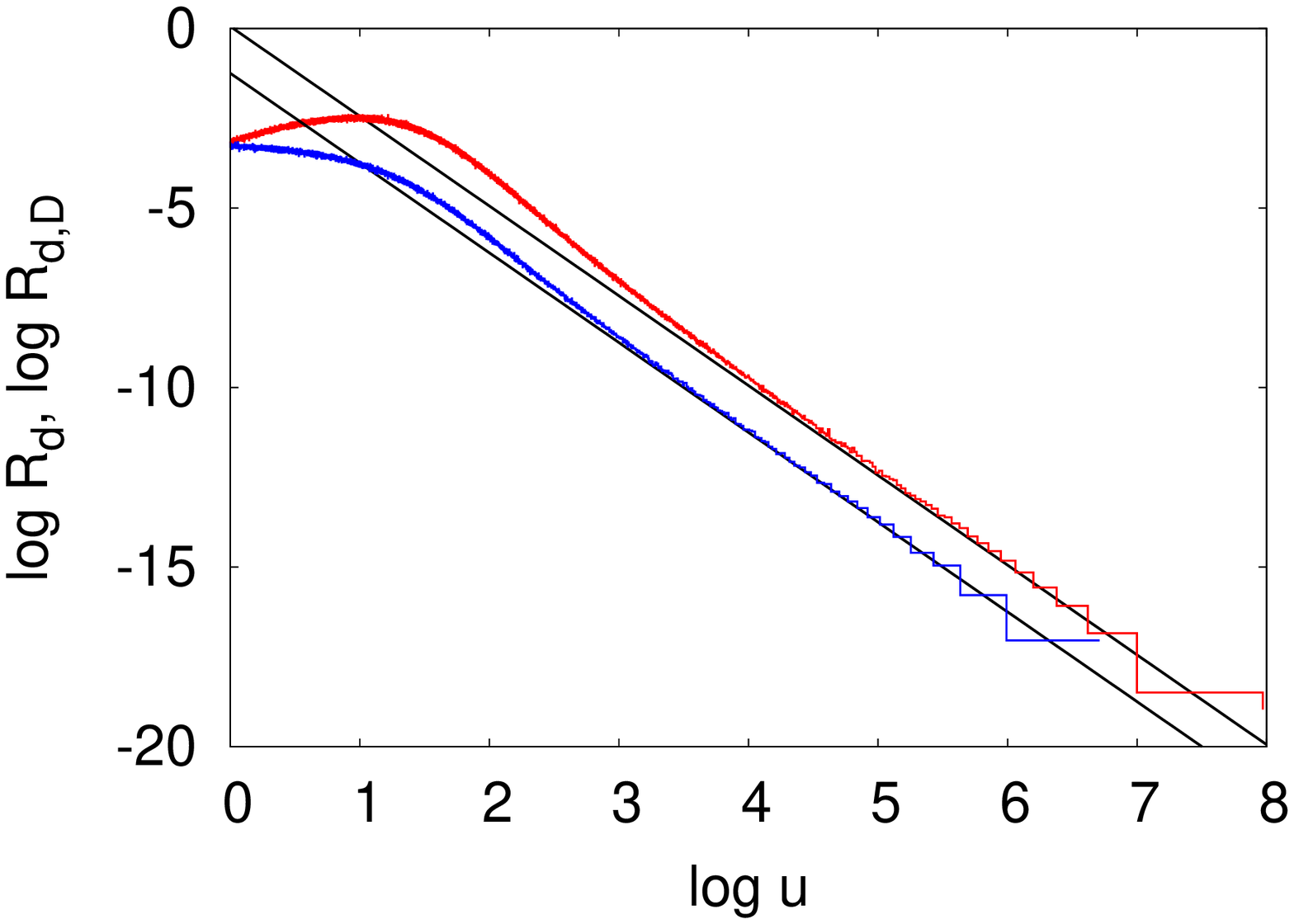} 
\caption{Two-dimensional L\'evy flights (\ref{2D_adim}) with $\alpha = 3/2$, $\gamma = 1$ and $c = 1$ and $n = 5000$ steps ($10^7$ samples). {\bf Left}: Survival probability in the wedged domain $q_{\cal D}(n)$ (red). A fit of the tail yields $q_{\cal D}(n) \sim n^{-\theta_{\cal D}}$ with $\theta_\mathcal{D} = 0.73 \pm 0.01$. {\bf Right}: Comparison of the PDF of the rescaled variable in the presence, $R_{d,{\cal D}}(u,0)$ (red), and in the absence, $R_d(u,0)$ (blue), of the absorbing wedge. The tails are in good agreement with our conjecture (\ref{R2d_specific}).}
\label{f:5:8}
\end{center}
\end{figure}

Here we consider the concrete example of a two-dimensional L\'evy random walker (see the left panel of Fig. \ref{f:0b:0d}). Its position $\vec{r}_n = x_n \,\vec{e_x} + z_n \,\vec{e_z}$ after $n$ steps evolves, for $n \geq 1$ according to  
\begin{eqnarray}\label{2D_walk}
\begin{cases}
x_{n} = x_{n-1} + \eta^x_{n} \;,\vspace{2mm}\\
z_{n} = z_{n-1} + \eta^z_{n} \;,
\end{cases}
\end{eqnarray}
starting from $\vec{r}_0 = \vec{0}$ at initial time. The jumps $ \eta^{x,z}_1,\eta^{x,z}_2,\ldots,\eta^{x,z}_n$ are independent and identical random variables, distributed according to the symmetric ($\gamma=1$) Pareto probability distribution $\phi(\eta)$ with $\alpha = 3/2$. We denote by $u$ and $v$ the rescaled variables:
\begin{eqnarray}\label{2D_adim}
u = \frac{x_n}{n^{1/\alpha}} \qquad, \qquad
v = \frac{z_n}{n^{1/\alpha}}\;, \; {\rm with} \; \; \vec y = (u,v) \;.
\end{eqnarray}
In absence of boundaries, the PDF of the rescaled variable $\vec y$ is easily obtained as the two components $x_n$ and $z_n$ are two independent one-dimensional L\'evy flights:
\begin{eqnarray}\label{R2d}
R_d(\vec y) = R(u) R(v) \;,
\end{eqnarray} 
where $R(u)$ is an $\alpha$-stable distribution (\ref{free_levy}) with $\beta = 0$ in the present case. We consider as open domain ${\cal D}$ the wedge depicted in the right panel of Fig.~\ref{f:0b:0d} and defined by $-\pi/3 \leq {\rm atan}(z/x) \leq \pi/3$. The fraction of walks which stay inside ${\cal D}$ after $n$ steps defines the survival probability $q_{\cal D}(n)$ which 
we compute numerically (see the left panel of Fig. ~\ref{f:5:8}). The persistence exponent extracted from our data is $\theta_\mathcal{D} = 0.73 \pm 0.01$.   

In this geometry, our result (\ref{conjecture_nD_section}) implies in particular that
\begin{eqnarray}\label{R2d_specific}
\frac{R_{d,{\cal D}}(u,v=0)}{R_{d}(u,v=0)} \underset{u\to \infty}{\longrightarrow}  \frac{1}{1-\theta_{\cal D}} \;.
\end{eqnarray}
In practice, we compute these quantities $R_{d}(u,v=0), R_{d,{\cal D}}(u,v=0)$ via $N_{\epsilon}(u)$, i.e. the number of points inside the rectangle $[u,u+\Delta u] \times [-\epsilon/n^{1/\alpha},\epsilon/n^{1/\alpha}]$, with $\epsilon$ and $\Delta u$ small (see the right panel of Fig. \ref{f:0b:0d}). In the absence of boundaries, it is easy to see that 
\begin{eqnarray}\label{numerical}
R_d(u,0) = \lim_{\epsilon, \Delta u \to 0} \frac{N_{\epsilon}(u)}{2 \epsilon \Delta u} = R(0) R(u) \;.
\end{eqnarray}
For large $u$ it behaves like
\begin{eqnarray}
R_d(u,0) \sim \frac{c R(0)}{u^{1+\alpha}} \;, \; R(0) = \frac{\Gamma(1+\alpha^{-1})}{a \pi} \;,
\end{eqnarray}
where $a$ and $c$ are related via Eq.~(\ref{c_and_gamma}). In particular, in our simulations with $\alpha = 3/2$ and $c=1$ we have  
\begin{eqnarray}\label{def_c*}
R_d(u,0) \sim \frac{c^*}{u^{5/2}} \;, \; c^* = \left(\frac{4 \sqrt{2\pi}}{3} \right)^{2/3} \;.
\end{eqnarray}
This relation has been checked numerically, as shown in the right panel of Fig. \ref{f:5:8}. Repeating the same numerical procedure in presence of the edge, we obtain $R_{d,{\cal D}}(u,0)$ as shown in the right panel of Fig. \ref{f:5:8}. The tail is in good agreement with our prediction $R_{d,{\cal D}}(u,0) \sim [c^*/(1-\theta_{\cal D})] u^{-1-\alpha}$, which confirms the validity of our conjecture (\ref{conjecture_nD_section}) in two dimensions.

\section{Conclusion}
To conclude, we have studied, in this paper, the problem of asymmetric L\'evy flights in presence of absorbing boundaries. {In the one dimensional case we gave a detailed derivation of the persistence exponents $\theta_+$ and $\theta_-$ for walkers constrained to stay in the semi-positive or semi-negative axis.} These exponents are useful for instance to characterize the statistical behavior of various observables including, for instance, the sequence of records for the walker position \cite{Wergen}. {Our main results concern the statistics of the walker position in a semi-bounded domain}. Far from the boundaries the PDF has the same algebraic decay as the original jump distribution: here we have computed with heuristic arguments and numerical simulations the amplitude of this decay. This last result strongly relies on the property that the statistics of this random walk is dominated by rare and large events and thus does not hold for the more familiar Brownian walkers.

\ack
G. S. acknowledges support by ANR grant 2011-BS04-013-01 WALKMAT. This project was also supported, in part, by the Indo-French 
Centre for the Promotion of Advanced Research under Project~$4604-3$.



\section*{References}

\begin{thebibliography}{100}

\bibitem{pareto}
V.~Pareto, {\em Cours d'\'economie politique}. Droz, Geneva (1896, 1965).

\bibitem{mandelbrot2}
B.~B. Mandelbrot, {\em Journal of Business} {\bf 36}, 394 (1963).

\bibitem{shlesinger_book}
e.~b. M.~F. Shlesinger, G.~M. Zaslavsky, U.~Frisch, {\em L\'evy Flights and
  Related Topics in Physics}. Springer, Berlin (1994).

\bibitem{biroli}
G.~Biroli, J.-P. Bouchaud, M.~Potters, {Europhys. Lett.} {\bf 78}, 10001
  (2007).

\bibitem{MSVV2013}
S. N. Majumdar, G. Schehr, D. Villamaina, P. Vivo, J. Phys. A: Math. Theor. {\bf 46}, 022001 (2013). 



\bibitem{bouchaud}
J.-P. Bouchaud, A.~Georges, {Phys. Rep.} {\bf 195}, 127 (1990).

\bibitem{mercadier}
N.~Mercadier, W.~Guerin, M.~Chevrollier, R.~Kaiser, {Nat. Phys.} {\bf 5},
  602 (2009).

\bibitem{koren1}
T.~Koren, A.~Chechkin, J.~Klafter, {Physica A} {\bf 379}, 10 (2007).

\bibitem{Stanley}
B.~Podobnik, A.~Valentin{\v c}i{\v c}, D.~Horvati\'c, H.~E. Stanley, {P.
  Natl. Acad. Sci. USA.} {\bf 108}(44), 17883–17888 (2011).

\bibitem{dario}
G. Gradenigo, A. Sarracino, D. Villamaina, T. S. Grigera, A. Puglisi, J. Stat. Mech. L12002 (2010).

\bibitem{feller_book}
W.~Feller, {\em An Introduction to Probability Theory and Its Applications}.
  Wiley, New York (1968).

\bibitem{hughes_book}
B.~D. Hughes, {\em Random Walks and Random Environments}, vol.~1. Clarendon
  Press, Oxford (1996).

\bibitem{metzler}
R.~Metzler, J.~Klafter, {Phys. Rep.} {\bf 339}, 1 (2000).

\bibitem{SatyaReview}
S.~N. Majumdar, {Curr. Sci.} {\bf 77}, 370 (1999).

\bibitem{Bray}
A.~J. Bray, S.~N. Majumdar, G.~Schehr, {preprint arxiv:1304.1195}  (2013).

\bibitem{klafter_image}
A. V. Chechkin, R. Metzler, V. Y. Gonchar, J. Klafter, L.
V. Tanatarov, J. Phys. A.: Math. Gen. {\bf 36}, L537 (2003).


\bibitem{BBDG}
A. Baldassarri, J.-P. Bouchaud, I. Dornic, C. Godr\`eche, Phys. Rev. E {\bf 59}, R20 (1999). 


\bibitem{koren2}
T.~Koren, M.~A. Lomholt, A.~V. Chechkin, J.~Klafter, R.~Metzler, {Phys.
  Rev. Lett.} {\bf 99}, 160602 (2007).

\bibitem{dybiec}
B. Dybiec, E. Gudowska-Nowak, P. H\"anggi, Phys. Rev. E {\bf 75}, 021109 (2007). 

\bibitem{Zumofen_Klafter}
G.~Zumofen, J.~Klafter, {Phys.~Rev.~E} {\bf 51}, 2805 (1995).

\bibitem{Rosso_Schehr}
R.~Garc\'ia-Garc\'ia, A.~Rosso, G.~Schehr, {Phys. Rev. E} {\bf 86}, 011101
  (2012).

\bibitem{zoia_rosso}
A.~Zoia, A.~Rosso, M.~Kardar, {Phys. Rev. E} {\bf 76}, 021116 (2007).

\bibitem{WMS12}
G.~Wergen, S.~N. Majumdar, G.~Schehr, {Phys. Rev. E} {\bf 86}, 011119
  (2012).

\bibitem{SA54}
E.~Sparre~Andersen, {Math. Scand.} {\bf 2}, 195 (1954).

\bibitem{zolotarev_book}
V.~Zolotarev, {\em One-dimensional stable distributions}, vol.~65. Amer. Math.
  Soc., Transl. of Math. Monographs (1962, RI (Transl. of the original 1983 in
  Russian)).

\bibitem{hanggi}
B.~Dybiec, E.~Gudowska-Nowak, P.~H\"anggi, {Phys. Rev. E} {\bf 73}, 046104
  (2006).

\bibitem{SiAu12}
T.~Simon, F.~Aurzada, {preprint arxiv:1203.6554}  (2012).

\bibitem{SA53}
E.~Sparre~Andersen, {Math. Scand.} {\bf 1}, 263 (1953).

\bibitem{Krauth}
W.~Krauth, {\em Statistical Mechanics: Algorithms and Computations}. Oxford
  University Press, Oxford (2006).



\bibitem{Wergen}
S.~N. Majumdar, G.~Schehr, G.~Wergen, {J. Phys. A} {\bf 45}, 355002 (2012).

\end{thebibliography}

\end{document}